\documentclass[pdftex]{PoS}

\def\simge{\mathrel{%
       \rlap{\raise 0.511ex \hbox{$>$}}{\lower 0.511ex \hbox{$\sim$}}}}
\def\simle{\mathrel{
       \rlap{\raise 0.511ex \hbox{$<$}}{\lower 0.511ex \hbox{$\sim$}}}}

\title{Determination of the endpoint of the first order deconfiniement phase transition in the heavy quark region of QCD}

\ShortTitle{Determination of the endpoint of the first order transition in the heavy quark region of QCD}

\author{
\speaker{Shinji Ejiri}
$^{1}$, 
Shota Itagaki$^{2}$, Ryo Iwami$^{3}$, Kazuyuki Kanaya$^{4,5}$, 
Masakiyo Kitazawa$^{6,7}$, Atsushi Kiyohara$^{6}$, Mizuki Shirogane$^{2}$, 
Yusuke Taniguchi$^{8}$, Takashi Umeda$^{9}$ 
(WHOT-QCD Collaboration)
\\ \\
$^1$Department of Physics, Niigata University, Niigata 950-2181, Japan\\
\ \ E-mail: ejiri@muse.sc.niigata-u.ac.jp \\
$^2$Graduate School of Science and Technology, Niigata University, Niigata 950-2181, Japan\\
$^3$Track Maintenance of Shinkansen, Rail Maintenance 1st Department, East Japan Railway Company Niigata Branch, Niigata, Niigata 950-0086, Japan\\
$^4$Tomonaga Center for the History of the Universe, University of Tsukuba, Tsukuba 305-8571, Japan\\
$^5$Faculty of Pure and Applied Sciences, University of Tsukuba, Tsukuba, Ibaraki 305-8571, Japan\\
$^6$Department of Physics, Osaka University, Toyonaka, Osaka 560-0043, Japan\\
$^7$J-PARC Branch, KEK Theory Center, Institute of Particle and Nuclear Studies,
KEK, 203-1, Shirakata, Tokai, Ibaraki, 319-1106, Japan\\
$^8$Center for Computational Sciences, University of Tsukuba, Tsukuba, Ibaraki 305-8571, Japan\\
$^{9}$Graduate School of Education, Hiroshima University, Higashihiroshima, Hiroshima 739-8524, Japan
}

\abstract{
We study the endpoint of the first order deconfinement phase transition of 2 and 2+1 flavor QCD in the heavy quark region.
We perform simulations of quenched QCD and apply the reweighting method to study the heavy quark region.
The quark determinant for the reweighting is evaluated by a hopping parameter expansion.
To reduce the overlap problem, we introduce an external source term of the Polyakov loop in the simulation. 
We study the location of critical point at which the first order phase transition changes to crossover by investigating the histogram of the Polyakov loop and applying the finite-size scaling analysis.
We estimate the truncation error of the hopping parameter expansion, and discuss the lattice spacing dependence and the spatial volume dependence in the result of the critical point.
}

\FullConference{37th International Symposium on Lattice Field Theory - Lattice2019\\
16-22 June 2019\\
Wuhan, China}

\begin{document}

\section{Introduction}
\label{sec:intro}

In 2+1 flavor QCD, there exist two first order phase transition regions in the quark mass parameter space.
When all the three quarks are massless, the chiral phase transition is first order.
Many studies have been made to determine the critical mass where the first order transition at small quark masses changes to a crossover.
The critical mass turned out to be close to the physical point and thus its quantitative determination is phenomenologically important. 
However, the continuum limit of the critical mass has not been obtained conclusively.
Another first order transition region locates around the heavy quark limit.
When all the quarks are infinitely heavy, QCD is just the pure gauge SU(3) Yang-Mills theory (quenched QCD), which is known to have a first order deconfinement transition.
This first order transition changes to crossover when the quark mass becomes smaller than a critical value.
However, its continuum limit is also not well understood.

In this study, we investigate the critical quark mass on lattices with $N_t=6$ and $8$, 
extending our previous study at $N_t=4$~\cite{Saito1,Saito2}.
We first use a histogram method combined with the reweighting method to study the critical point in 2 and 
2+1 flavor QCD in Sec.~\ref{sec:histogram}.
We also discuss the limitation of the reweighting from quenched QCD due to the overlap problem.
To reduce the overlap problem, we perform simulations of an effective action containing a Polyakov loop term in Sec.~\ref{sec:simplt}. 
We determine the critical point by the finite volume scaling analysis.
Section~\ref{sec:summary} is devoted to a summary.

\section{Histogram method}
\label{sec:histogram}

We study the phase structure of QCD around the endpoint of first order phase transition line by the histogram of the absolute value of the Polyakov loop $|\Omega|$, which we define as 
\begin{eqnarray}
W( |\Omega|; \beta, K) 
= \int {\cal D} U \ \delta(|\Omega| - |\hat{\Omega}|) \ 
e^{-S_g}\ \prod_{f=1}^{N_{\rm f}} \det M(K_f) , \ \ \ 
S_g = 6N_{\rm site} \beta \hat{P} ,
\label{eq:hist}
\end{eqnarray}
where $\hat{P}$ is the average plaquette, $\det M$ is the quark determinant, $\beta=6/g^2$ is the gauge coupling, $N_{\rm site}=N_s^3 \times N_t$ is the number of lattice sites, $N_{\rm f}$ is the number of flavors, and $K_f$ is the hopping parameter for the $f^{\rm th}$ flavor.
For the quarks, we adopt the standard Wilson quark action.
In terms of the histogram, the probability distribution function of $|\Omega|$ is given by 
${\cal Z}^{-1}(\beta, K) \, W(|\Omega|;\beta,K)$ 
with $\cal Z$ the partition function defined by
$ {\cal Z}(\beta,K)  = \int\! W(|\Omega|;\beta,K) \, d|\Omega| $.

We then define our effective potential by $V_{\rm eff} (|\Omega|) = - \ln W (|\Omega|)$.
On a first order transition line, $V_{\rm eff}$ is double-well type.
The double-well turns into a single-well when the quark mass closses the endpoint of the first order transition line which we call the critical point.

We adopt the hopping parameter expansion to compute the quark determinant in the heavy quark region \cite{Saito1}.
Up to the next-to-leading order contributions, the quark determinant for each flavor is expanded as 
\begin{eqnarray}
 \ln \det M(K) &=&
  288N_{\mathrm{site}}K^4 \hat{P} + 768N_{\mathrm{site}}K^6
  \left( 3 \hat{W}_{\mathrm{rec}}+6 \hat{W}_{\mathrm{chair}}+2 \hat{W}_{\mathrm{crown}} \right) + \cdots
\nonumber \\
  && \hspace{-15mm}
 +12\times 2^{N_t}N_s^3 K^{N_t} \mathrm{Re} \hat\Omega 
 + 36\times 2^{N_t}N_s^3 N_t K^{N_t+2} \left( 2 \! \sum_{n=1}^{N_t/2-1} \! \mathrm{Re} \hat\Omega_n  
 +  \mathrm{Re} \hat\Omega_{N_t/2} \right) + \cdots  \ ,
 \label{eq:hpe_nlo}
\end{eqnarray}
where $\hat{W}_{\mathrm{rec}}$, $\hat{W}_{\mathrm{chair}}$, $\hat{W}_{\mathrm{crown}}$ are the 6-step Wilson loop operators of rectangle-type, chair-type, and crown-type, respectively, 
and $\hat{\Omega}_n$ is the $(N_t+2)$-step bended Polyakov loop, which contains two spatial links and $n$ temporal links between these spatial links.
The first term with $\hat{P}$ can be absorbed into the plaquette gauge action by a shift $
\beta \rightarrow \beta^* \equiv \beta + 48 \sum_{f=1}^{N_{\rm f}} K_f^4$, where we have summed up contributions of all flavors. 
The terms with 6-step Wilson loops can also be absorbed into improvement terms of improved gauge action. 
Because a shift in improvement parameters only affects the amount of lattice discretization errors within the same universality class, the 6-step Wilson loop terms will not affect characteristic physical properties of the system in the continuum limit, such as the order of the phase transition.
Therefore, in this study, we concentrate on the influence of Polykov-loop-type terms in~Eq.~(\ref{eq:hpe_nlo}).
To study the truncation error of the hopping parameter expansion, we perform both the leading order (LO) calculation keeping only the $O(K^{N_t})$ term, and the next-to-leading order (NLO) calculation in which  the $O(K^{N_t+2})$ terms are also taken into account.

We investigate how the shape of $V_{\mathrm{eff}}$ is changed by variation of the quark mass.
We thus need to know $V_{\mathrm{eff}}$ in a wide range of $|\Omega|$. 
This is not straightforward with a single simulation because the statistical accuracy of $V_{\mathrm{eff}}$ quickly drops down off the minimum point where $|\Omega|$ is the most probable.
To confront the issue, we combine information at several simulation points by the multi-point reweighting method~\cite{Saito1,Saito2}. 
To stay close to the first order transition line, we adjust $\beta$ for each $K$ so that two minimum values of $V_{\mathrm{eff}}$ are as equal as possible.
We then measure the difference between the peak height in the middle of $V_{\mathrm{eff}}$ and the minimum value, which is denoted as $\Delta V_{\mathrm{eff}}$. 
We define the critical point $K_{ct}$ as $K$ where the difference $\Delta V_{\mathrm{eff}}$ vanishes.

We perform simulations of quenched QCD on $24^3 \times 6$, $32^3 \times 6$, $36^3 \times 6$, and $24^3 \times 8$ lattices using the pseudo heat bath algorithm with over relaxation.
We investigate the critical mass mainly on the $N_t =6$ lattices. 
The lattice spacing $a$ is given by $a=(N_t T_c)^{-1}$ for a simulation at the transition temperature $T_c$.
To evaluate the lattice spacing dependence, i.e., $N_t$ dependence, we use the results for $N_t=4$ obtained in Ref.~\cite{Saito1} and perform an additional simulation on a lattice with $N_t=8$.  
Simulations at 4-6 $\beta$ points around the transition point are combined by the multi-point histogram method.
Details of the simulations are given in~Ref.~\cite{itagaki19}.

\subsection{Critical point in two flavor QCD}

\subsubsection{Results at $N_t=6$}

\begin{figure}[tb]
\centering
\vspace{-8mm}
\includegraphics[width=7cm,clip]{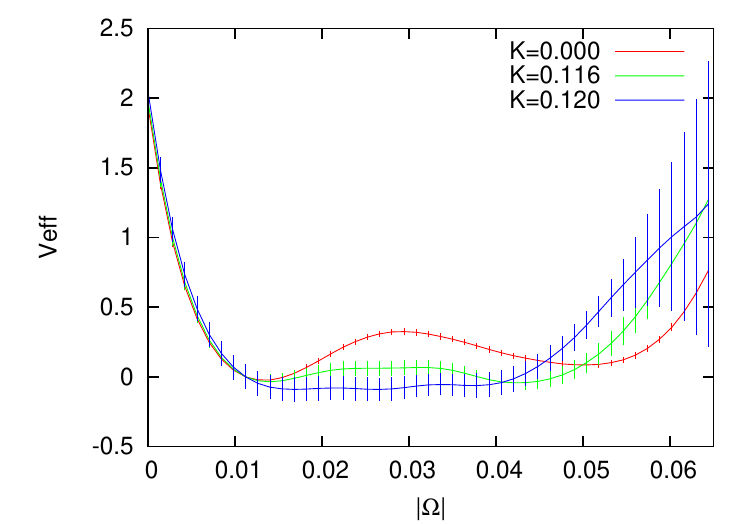}
\hspace{1mm}
\includegraphics[width=7cm,clip]{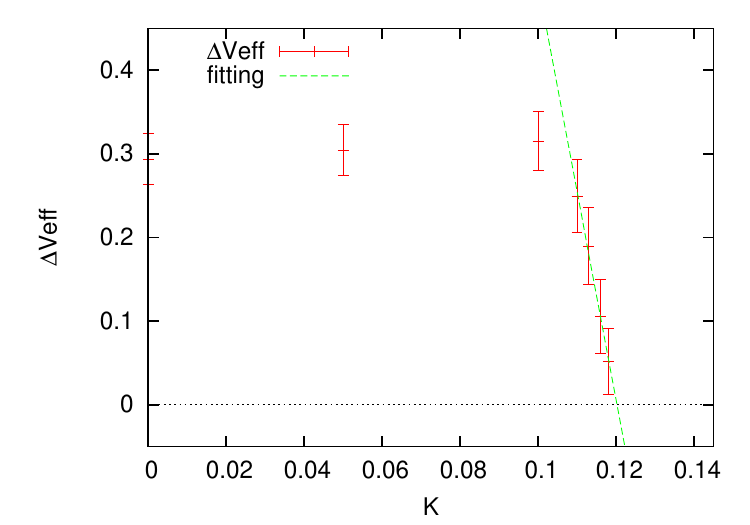}
\vspace{-3mm}
\caption{$V_{\mathrm{eff}}$ (left) and $\Delta V_{\mathrm{eff}}$ (right) obtained by the hopping parameter expansion up to the next to leading order on a $24^3\times6$ lattice.
}
\label{fig1}
\end{figure}

In the left panel of Fig.~\ref{fig1}, we show $V_{\mathrm{eff}}$ for two flavor QCD on the $24^3 \times 6$ lattice computed with the NLO $\ln \det M$ given in~Eq.~(\ref{eq:hpe_nlo}).
We find two minima for $K=0.0$ and 0.116, while it becomes almost flat at the minimum for $K=0.120$.
We plot $\Delta V_{\mathrm{eff}}$ as a function of $K$ in the right panel of~Fig.~\ref{fig1}.
Fitting the smallest four data by a linear function (dashed line), we obtain $K_{ct}=0.1202(19)$.

Repeating the same calculation only with the LO term, we obtain $K_{ct}=0.1359(30)$ on the $24^3 \times 6$ lattice, and $K_{ct}=0.1286(40)$ on the $32^3 \times 6$ lattice.
Though these $K_{ct}$ with different spatial volumes are roughly the same within the statistical errors, 
their central values may be suggesting that $K_{ct}$ decreases at $N_t=6$ as the spatial volume increases.
We have also tried to calculate $K_{ct}$ on a $36^3 \times 6$ lattice. 
However, unlike the cases of $24^3 \times 6$ and $32^3 \times 6$ lattices, the overlap problem turned out to be severe on this lattice to obtain a reliable $V_{\mathrm{eff}}$ up to the critical point~\cite{itagaki19}.
We come back to the issue of overlap problem in Sec.~\ref{sec:simplt}.

\paragraph{Effective NLO method:}
Before proceeding to other issues, let us discuss a method, introduced in~Ref.~\cite{Saito2}, to effectively incorporate NLO effects in the LO calculation of $V_{\mathrm{eff}}$ and $K_{ct}$.
The basic observation of~Ref.~\cite{Saito2} is that the bended Polyakov loops $\hat{\Omega}_n$ have strong linear correlation with $\hat{\Omega}$ on each configuration and are well approximated by 
${\rm Re} \hat\Omega_n \approx c_n \,{\rm Re} \hat\Omega$, where $c_n = \langle {\rm Re} \hat\Omega_n / {\rm Re} \hat\Omega \rangle$.
Substituting this into Eq.~(\ref{eq:hpe_nlo}), we find that the NLO effects can be absorbed by a shift of $K$.
Denoting $K_{ct}$ calculated only with the LO term as $K_{ct,\mathrm{LO}}$, the $K_{ct}$ effectively including the NLO terms can be obtained by solving
\begin{equation}
K_{ct,{\rm eff}}^{N_t} \left( 1+C_\Omega\,N_t K_{ct,{\rm eff}}^{2} \right)=  K_{ct,\mathrm{LO}}^{N_t}, \hspace{8mm}
C_\Omega \equiv 6\sum_{n=1}^{N_t/2-1} c_n +3c_{N_t/2}.
\label{eq:keff}
\end{equation}
On the $24^3 \times 6$ lattice, we find that $K_{ct,{\rm eff}}=0.1205(23)$, which is consistent with $K_{ct}=0.1202(19)$ computed directly with the NLO contributions.
We thus find that the effective NLO method works well.
The method is useful in avoiding repeated analyses, e.g., for various number of flavors.

\paragraph{Truncation error of hopping parameter expansion:}
We now compare the LO and NLO results for $K_{ct}$.
We find that NLO $K_{ct}$ is about 10\% smaller than the LO results.
This means that the truncation error of the hopping parameter expansion in $K_{ct}$ is larger than the statistical errors for $N_t=6$. 
This is in contrast to the case of $N_t=4$:
On the $24^3 \times 4$ lattice, we obtain $K_{ct,{\rm eff}}=0.0640(10)$ with the effective NLO method, to be compared with the LO value $K_{ct,\mathrm{LO}}=0.0658(3)(^{+4}_{-11})$~\cite{Saito1}. 
We thus find that, for $N_t=4$, the truncation error is about 3\% and small in comparison with the statistical errors~\cite{Saito2}.
Careful treatments including higher order terms are required for $N_t \ge 6$.


\subsubsection{Towards the continuum limit of the critical point}
Our results of $K_{ct}$ on $24^3 \times 4$ and $24^3 \times 6$ lattices are summarized in Table~\ref{tab1}.
The effective NLO method was used for the NLO value for $N_t=4$.
%
We note that $K_{ct}$ for $N_t=6$ is about twice larger than that for $N_t=4$. 
We have also calculated $V_{\mathrm{eff}}$ on a $24^4 \times 8$ lattice.
However, $V_{\mathrm{eff}}$ remains double-well up to quite large $K$ where the hopping parameter expansion is not applicable. 
Thus, we can not find $K_{ct}$ for $N_t=8$ with the present method~\cite{itagaki19}.

\begin{table}[tb]
\vspace{-8mm}
  \caption{
  The critical point $K_{ct}$ on $24^3 \times 4$ and $24^3 \times 6$ lattices calculated by LO and NLO hopping parameter expansion of two flavor QCD. Also listed are the values of $m_{\mathrm{PS}}/T_c$ at $K_{ct}$,  computed by the reweighting method and by the two flavor full QCD simulation, on a zero-temperature $16^3 \times 32$ lattice.}
 \begin{center}
  \begin{tabular}{c|ccc|ccc}
   \hline
   & up to LO & reweighting     & full QCD  & up to NLO & reweighting     & full QCD \\ 
   lattice         & $K_{ct}$ & $m_{\mathrm{PS}}/T_c$ &  $m_{\mathrm{PS}}/T_c$  & $K_{ct}$ & $m_{\mathrm{PS}}/T_c$ & $m_{\mathrm{PS}}/T_c$ \\ \hline
   $24^3 \times 4$ & 0.0658(10) &  15.47(14) &  15.47(14) & 0.0640(10) &  15.74(14) &  15.73(14) \\ 
   $24^3 \times 6$ & 0.1359(30) &   7.88(69) &  7.43(78)   & 0.1202(19) &  11.29(40) &  11.15(42) \\ \hline
  \end{tabular}
  \label{tab1}
 \end{center}
\end{table}


To make the physical implication of the values of $K_{ct}$ clearer, we calculate the pseudo-scalar meson mass $m_{\mathrm{PS}}$ at $K_{ct}$ by performing additional zero-temperature simulations.
In this study, we perform the following two simulations:
One is the direct two flavor full QCD simulation adopting the same combination of gauge and quark actions and adjusting the simulation parameters $(\beta, K)$ to the critical point obtained in the finite-temperature study.
Another is the quenched QCD simulation combined with the reweighting method, as adopted in the determination of $K_{ct}$ with the LO hopping parameter expansion.
As we discussed, the effect of the plaquette term can be absorbed by the shift $\beta \rightarrow \beta^*$. 
In both full and quenched simulations, we generate configurations by the hybrid Monte Carlo algorithm on $16^3 \times 32$ lattices. 
The number of configurations is 52 at each simulation point. 

Our results of $m_{\mathrm{PS}}/T_c$ at $K_{ct}$ are listed also in Table~\ref{tab1}.
The errors for $m_{\mathrm{PS}}/T_c$ contain that propagated from the error of $K_{ct}$.
We find that the results of full QCD and reweighting calculations are consistent within the errors.
This means that the reweighting method is effective and the disregarded 6-step Wilson loops in the reweighting method have no large effects also on the pseudo-scalar meson mass.
On the other hand, for $N_t=6$, corresponding to the difference of $K_{ct}$ between the LO and NLO calculations, 
$m_{\mathrm{PS}}/T_c$ at the NLO $K_{ct}$ is $1.4$ times smaller than that at the LO $K_{ct}$.
If we estimate the systematic error from the truncation of the hopping parameter expansion by the difference between the LO and NLO calculations, we obtain $m_{\mathrm{PS}}/T_c = 11.15(42)(372)$ for $N_t=6$.
Because this is smaller than $m_{\mathrm{PS}}/T_c = 15.73(14)(26)$ for $N_t=4$, 
our results suggest that the critical quark mass decreases as the lattice spacing decreases.

\subsection{Critical line in 2+1 flavor QCD}
\label{sec:3flavor}

\begin{figure}[tb]
\centering
\vspace{-8mm}
\includegraphics[width=7cm,clip]{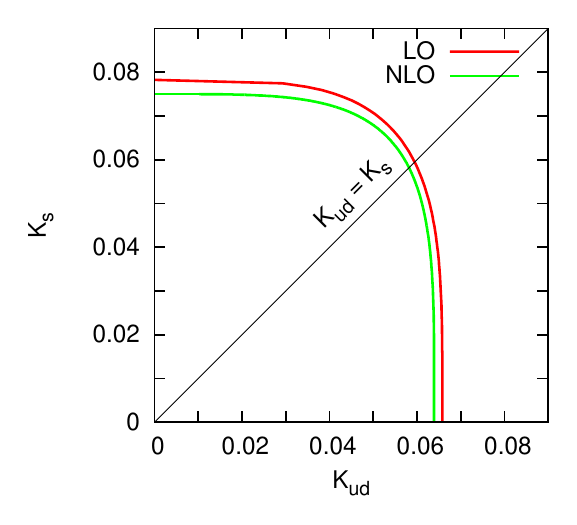}
\hspace{1mm}
\includegraphics[width=7cm,clip]{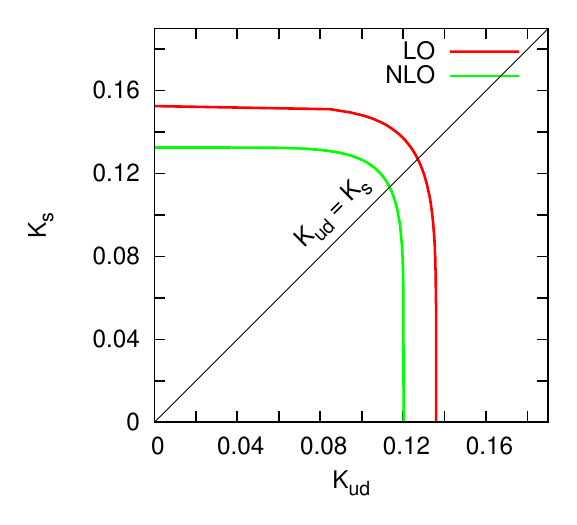}
\vspace{-5mm}
\caption{Critical line of 2+1 flavor QCD on the $24^3 \times 4$ (left) and $24^3 \times 6$ (right) lattices.
}
\label{fig2}
\end{figure}

The effective NLO relation~(\ref{eq:keff}) can be easily generalized to any number of flavors. 
E.g., the critical line in the $(K_{ud},K_s)$ space in 2+1 flavor QCD is given by 
\begin{eqnarray}
 2 K_{ct,ud}^{N_t}  \left( 1+C_\Omega\,N_t K_{ct,ud}^{2} \right)
 + K_{ct,s}^{N_t}   \left( 1+C_\Omega\,N_t K_{ct,s}^{2} \right)
 =2 K_{ct,\mathrm{LO}}^{N_t}  ,
\label{eq:2+1clNLO}
\end{eqnarray}
where
$K_{ct,\mathrm{LO}}$ is the LO $K_{ct}$ for $N_{\rm f}=2$. 
By solving this equation using $K_{ct,\mathrm{LO}}$ obtained on the $24^3 \times 4$ and $24^3 \times 6$ lattices, 
we obtain green curves in Fig.~\ref{fig2} for $N_t=4$ (left) and $N_t=6$ (right), respectively.
The red curves in Fig.~\ref{fig2} are for the LO critical line calculated by Eq.~(\ref{eq:2+1clNLO}) with the terms proportional to $C_\Omega$ removed.
The difference between the two curves is an estimate for the truncation error of the hopping parameter expansion.

\section{Effective heavy quark QCD with Polyakov loop}
\label{sec:simplt}

\begin{figure}[tb]
\centering
\includegraphics[width=7cm,clip]{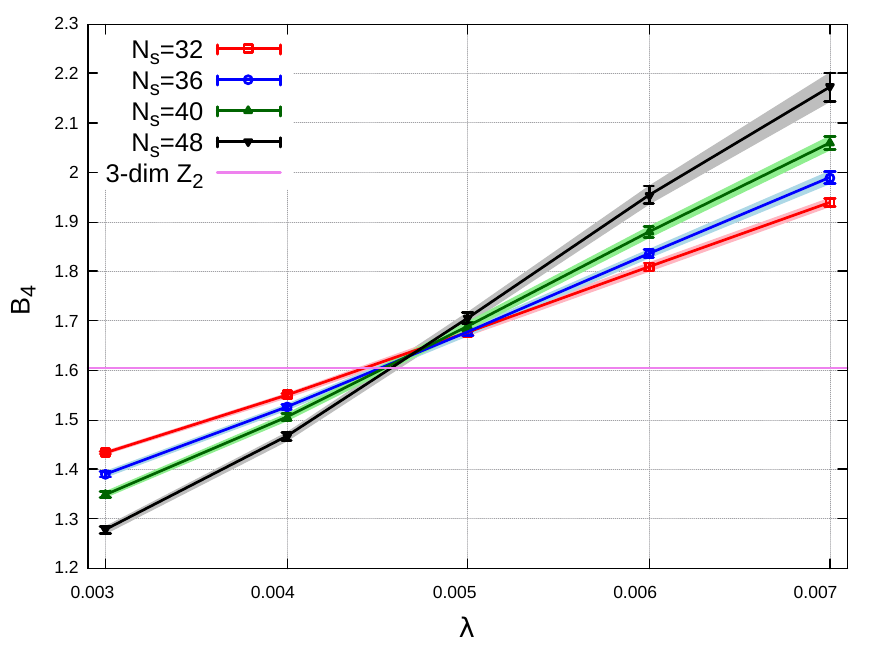}
\vspace{-3mm}
\caption{$\lambda$ dependence of $B_4$ at the transition point.
}
\label{fig3}
\end{figure}

We found that the overlap problem occurs in the determination of $K_{ct}$ on the $36^3 \times 6$ lattice~\cite{itagaki19}.
As an attempt to avoid the overlap problem, we perform simulations with an effective action for QCD with heavy quarks, $S_{\rm eff} = -6 N_{\rm site} \beta \hat{P} - N_s^3 \lambda {\rm Re} \hat{\Omega}$.
The Polyakov loop term corresponds to the LO hopping parameter expansion of $\ln \det M$ with $\lambda = 12 \times 2^{N_t} N_{\rm f}  K^{N_t}$.
Because a heat bath algorithm is applicable to this action, the computational cost is much smaller than full QCD simulations.
We include the NLO contributions by reweighting.

Simulations are performed on $N_t=4$ lattices with $N_s=32$, 36, 40, and 48.
We generate gauge configurations for several values of $(\beta,\lambda)$, and study dependence on these parameters by the multi-point reweighting method.
The number of configurations is $600,000$ for each $\beta$ and $\lambda$.
In this study, we identify the critical point by the Binder cumulant of the Polyakov loop,
\begin{eqnarray}
B_4=\frac{\langle ( \Omega - \langle \Omega \rangle )^4 \rangle}{\langle ( \Omega - \langle \Omega \rangle )^2 \rangle^2 }.
\end{eqnarray}
At the critical point $\lambda_{ct}$, $B_4$ is independent of the spatial volume.
The value of $B_4$ at $\lambda_{ct}$ depends on the universality class.

The results of $B_4$ on the first order transition line are plotted in Fig.~\ref{fig3} as function of $\lambda$.
As shown in this figure, the lines of $B_4$ with different volumes cross at one point.
We fit the data with 
$B_4(N_s, \lambda)=B_{4ct}+AN_s^{1/\nu}(\lambda -\lambda_{ct})$,
where $B_{4ct}$, $A$, $\nu$ and $\lambda_{ct}$ are the fit parameters.
$B_{4ct}$ and $\lambda_{ct}$ correspond to $B_4$ and $\lambda$ at the critical point, respectively.
From fit using data of all four volumes, we obtain $\lambda_{ct} =0.004754(84)$, 
$B_{4ct} =1.644(13)$ and $\nu =0.65(8)$.
When we fit data of the largest three volumes corresponding to $N_s =36$, 40, and 48 only,
we get $\lambda_{ct}=0.00468(11)$, $B_{4ct} =1.630(20)$ and
$\nu =0.65(11)$.
These results are almost consistent with those expected from the universality class of the 3D Ising model: $B_{4ct}^\mathrm{Ising}=1.604$ and $\nu^\mathrm{Ising}=0.63$.

\begin{figure}[tb]
\centering
\vspace{-8mm}
\includegraphics[width=65mm,clip]{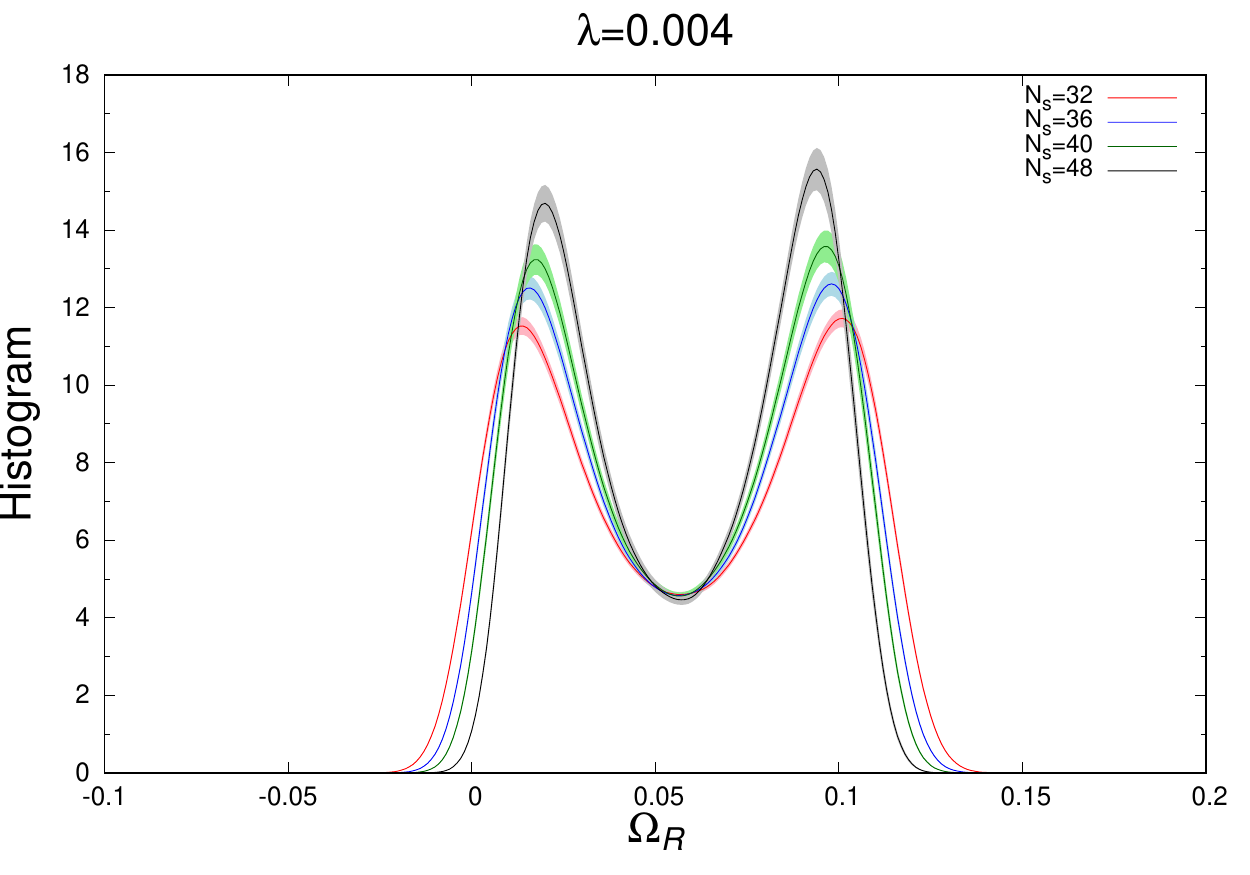}
\hspace{1mm}
\includegraphics[width=65mm,clip]{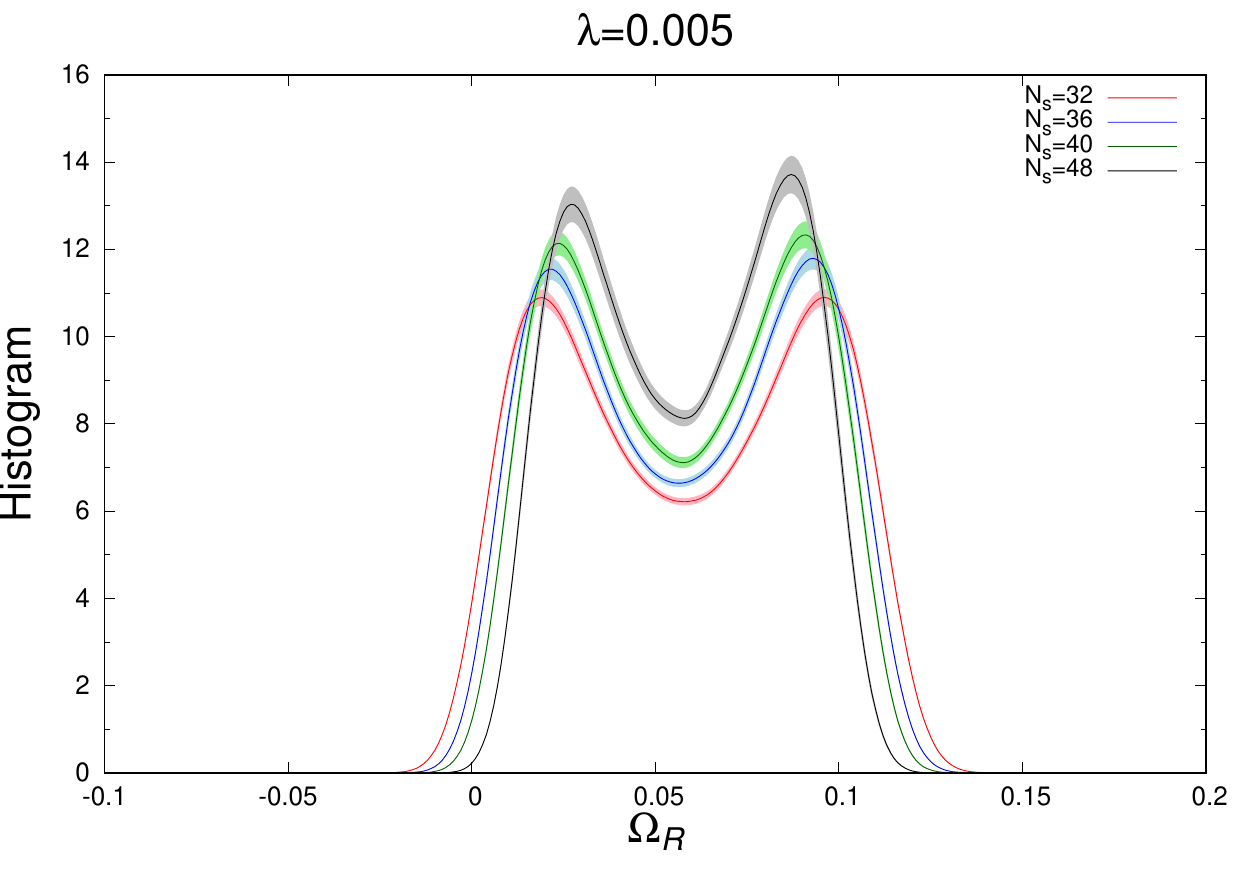}
\vspace{-3mm}
\caption{Spatial volume dependence of the histograms at $\lambda =0.004$ and $0.005$.}
\label{fig4}
\end{figure}

The result $\lambda_{ct} =0.004754(84)$ corresponds to $K_{ct}=0.05932$.
This is $10\%$ smaller than the result obtained by the histogram method on the $24^3 \times 4$ lattice.
To understand this difference, we investigate the volume dependence of the histogram at the transition point. 
As we mentioned in the previous section, $K_{ct}$ decreases as the volume increases.
The histograms just above and below $\lambda_{ct}$ are plotted in Fig.~\ref{fig4}.
For the case of $\lambda =0.004 < \lambda_{ct}$ (left), the central dent in the histogram gets deeper as the volume increases.
On the other hand, for $\lambda =0.005 > \lambda_{ct}$ (right), the central dent becomes shallower.
These are consistent with the picture that 
$\lambda_{ct}$ is the boundary that divides regions with one peak and two peaks in the volume infinity limit.

\section{Summary}
\label{sec:summary}

We studied the location of critical point at which the first order phase transition changes to crossover in the heavy quark region by investigating the histogram of the Polyakov loop and applying the finite-size scaling analysis.
We performed simulations of quenched QCD together with reweighting method.
The quark determinant is evaluated by the hopping parameter expansion.
Truncation error of the hopping parameter expansion is visible for $N_t=6$.
Higher order terms are needed for large $N_t$.
Overlap problem arises for large volume.
To reduce the overlap problem, we introduce an external source term of the Polyakov loop in the simulation.
The value of $B_4$ at the critical point is almost consistent with that of 3D ising model.

\paragraph{Acknowledgments:} 
This work was in part supported by JSPS KAKENHI (Grant Nos.\ 
JP19H05146, JP19K03819, JP19H05598, JP18K03607, JP17K05442, JP15K05041, JP26400251, and JP26287040), 
the HPCI System Research project (Project ID: hp170208, hp190028, hp190036), and JHPCN projects (jh190003, jh190063).
This research used 
OCTPUS at the Osaka University
and ITO at the Kyushu University.

\end{document}